\documentstyle[epsfig]{l-aa}

\newcommand{\etal}{{\em et al.}}

\newcommand{\mic}{\mbox{$\mu$m}}

\begin{document} 

%\baselineskip=0.6truecm

%\thesaurus{06; % A&A Section 6: Form. struct. and evolut. of stars
%           07;    % Stars: structure of.            
%} 
\thesaurus{11(11.09.1; 11.09.4; 11.19.2; 13.09.3; 08.06.3)}

\title{The dark side of star formation in the Antennae galaxies}
 
%\subtitle{............}

\author{I.F. Mirabel\inst{1}, L. Vigroux\inst{1}, V. Charmandaris\inst{1}, M. Sauvage\inst{1}, P. Gallais\inst{1}, D. Tran\inst{1}, C. Cesarsky \inst{1}, 
S.C. Madden\inst{1} \& P.-A. Duc\inst{2}}

\institute{CEA/DSM/DAPNIA Service d'Astrophysique F-91191 Gif-sur-Yvette, France
\and 
ESO, Karl-Schwarzschild-Str. 2, D-85748 Garching bei M\"unchen, Germany}

\date{Received 19 December 1997 / Accepted 3 February 1998}

\maketitle

\offprints{I.F. Mirabel} 

\markboth{Mirabel \etal\\: The dark side of star formation in the Antennae}{The dark side of star formation in the Antennae}

%----------------------------------------------------------------------------

\begin{abstract}
 
We compare mid-infrared images of the Antennae galaxies (NGC\,4038/39)
from the Infrared Space Observatory, with optical images from the
Hubble Space Telescope. The mid-infrared observations show that the
most intense starburst in this colliding system of galaxies takes
place in an off-nucleus region that is inconspicuous at optical
wavelengths. The analyses of the mid-infrared spectra indicate that
the most massive stars are being formed in an optically obscured knot
of 50 pc radius, which produces about 15\% of the total luminosity
from the Antennae galaxies between 12.5 \mic\ and 18 \mic. The
mid-infrared observations reported here demonstrate that the
interpretation of star formation properties in colliding/merging
systems based on visible wavelengths alone can be profoundly biased
due to dust obscuration. The multiwavelength view of this nearby
prototype merging system suggests caution in deriving scenarios of
early evolution of high redshift galaxies using only observations in
the narrow rest-frame ultraviolet wavelength range.


\end{abstract}

%----------------------------------------------------------------------------

\section{Introduction}

One of the most important recent discoveries in extragalactic
astronomy has been the identification of a class of ``infrared
luminous galaxies" (L$_{bol}$ $\geq$ 10$^{11}$ L$_{\odot}$), which
emit more energy in the infrared (5-500 $\mu$m) than in all other
wavelengths combined (see review by Sanders \& Mirabel, 1996).  The
trigger of the intense infrared emission appears to be the
interaction/merger of molecular gas-rich spirals.  Although the
spectrum of the integrated radiation from these galaxies suggests that
the bulk of the luminosity arises in regions that are heavily obscured
by dust, the actual distribution of the mid-infrared emission with
high spatial resolution is poorly known.

The Infrared Space Observatory (ISO, Kessler \etal, 1996) offers
unprecedented capabilities with respect to the Infrared Astronomical
Satellite (IRAS). In the mid-infrared (5.5$\mu$m - 16.5 $\mu$m) the
Infrared Space Observatory Camera (ISOCAM) provides an improvement in
sensitivity of $\sim$ 1000, and an increase in spatial resolution by a
factor of $\sim$ 60. Furthermore, observations with arcsec resolution
in spectrophotometric mode with Circular Variable Filters (CVFs)
permit us to infer the nature of the optically invisible stars that
heat the dust and ionize the gas.
 
Here we present for the first time an image of NGC\,4038/39 (Arp\,244
= VV\,245 = `The Antennae') in the LW3 (12 -17 \mic) filter with
1.5$''$ pixel field of view and compare it with the HST optical
image. An extensive account of all the ISOCAM observations will be
given in a forthcoming paper by Vigroux et al (1998). The Antennae is
a prototype nearby merger system of two late-type spiral disk galaxies
with nuclei separated by $\sim$ 6.4 kpc.  At a distance of 20 Mpc the
total infrared luminosity measured by IRAS is 10$^{11}$ L$_{\odot}$,
which is about five times the luminosity of the system at visual
wavelengths.  Molecular gas observations with a resolution of 6$''$
(Stanford \etal\ 1990) showed that $\sim$ 60\% of the CO(1-0) emission
from the overall system (Sanders \& Mirabel 1985) is concentrated in
the two nuclei and in an extended off-nuclear complex where the two
disks overlap. The spatial distribution of the CO emission is
consistent with the $\lambda$6 cm and $\lambda$20 cm radio continuum
maps by Hummel \& van der Hulst (1986).

\section{Observations and Data Analysis}

First observations of the Antennae with coarser resolution were
obtained during the Performance Verification phase (Vigroux et
al. 1996). The new ISOCAM results presented in this paper consist of a
large raster map obtained with the LW3 (12 -17 \mic) filter using
1.5$''$ pixel field of view, with a full width at half max of the
observed point spread function of 4.5$''$. Full CVF scans were also
made from 5.5 \mic\ to 16.5 \mic\, using the smallest possible
increment, 0.1 \mic\ step and 6$''$ pixel field of view (Vigroux et
al. 1996).

The data analysis was performed using the CAM Interactive Analysis
(CIA)\footnote{CIA is a joint development by the ESA astrophysics
division and the ISOCAM consortium led by the ISOCAM PI,
C. C\'{e}sarsky, Direction des Sciences de la Mati\`{e}re, C.E.A.,
France} Software. For the raster map, dark subtraction was performed
using a model of the secular evolution of ISOCAM's dark current
(Biviano et al., 1997). Removal of cosmic ray induced spikes was
performed using a multi-resolution median filtering technique (Stark
et al. 1996). The data cube was then corrected for the memory effect
of the LW detector using the inversion algorithm described in Abergel
et al. (1996). As the raster was rather large, 5$\times$5 pointings,
the flat-field was built from the median off the sky positions of the
galaxies.

\section{Comparison of the mid-infrared and optical images}

The LW3 flux (12-17 \mic) is shown in Figure 1 in red contours
superimposed on the HST WFPC2 combined images in the V (5252 \AA) and
I (8269 \AA) filters by Whitmore et al. (1997). The infrared emission
appears to be associated with the nuclei of both galaxies, the star
forming ring in the northern galaxy NGC 4038, and with the relatively
obscured overlap region of the two disks, which is $\sim$ 5x3 kpc in
extent. The emission is very clumpy and bright knots dominate the
mid-infrared emission, which as discussed below, comes from gas and
dust heated by massive stars.

The two optical nuclei are detected as bright knots in the
mid-infrared, but the brightest knot in the 15 \mic\ map is $\sim$ 2.3
kpc east of the nucleus of the southern galaxy NGC 4039. This knot is
toward the southern corner of the region of overlap of the two
disks. In this region, several optical red objects were found
(Whitmore \& Schweizer, 1995) associated with discrete intensity peaks
in the centimeter continuum maps (Hummel \& van der Hulst, 1986) and
millimeter CO maps (Stanford et al. 1990). The 5x3 kpc overlap region
contributes about half of the 2.16 Jy flux radiated from the whole
system at 15 $\mu$m. An average extinction of A$_{v}$ $\sim$ 70 mag in
the overlap region has been derived by Kunze \etal\ (1996) using SWS
ISO observations. A mean absorption of A$_{v}$ $\sim$ 73 mag was
independently inferred by Stanford \etal\ (1990) from CO
interferometric observations.

At 15 \mic\ the brightest area of the overlap region, located east of
the nucleus of NGC 4039, is resolved into two peaks separated by
6$''$. The fainter one has a bright WFPC2 visual counterpart. However,
the brighter eastern peak in the 15 \mic\ map has only a faint I band
counterpart of $\sim$ 48 pc in effective radius (Whitmore \&
Schweizer, 1995), but no conspicuous V band counterpart is seen in the
HST image of Figure 1. The brighter eastern knot alone contributes
$\sim$15\% of the total luminosity at 15 \mic\, and at this wavelength
it is almost 100 times brighter than the nearby nucleus of NGC 4039
(Vigroux et al. 1996).  From J, H, and K band images (Duc \& Mirabel,
1997), we measure an excess of A$_v$ $\sim$ 7 mag in the visual
absorption in front of the eastern knot relative to the western knot.

The most luminous knots at 15 \mic\ do not coincide with the most
prominent dark lanes of the HST image.  An even more striking
displacement between the mid-infrared and optical dust lanes has been
found in Centaurus A (Mirabel et al. 1998). These displacements could
be due to different spatial distributions of the warm dust that
radiates at 15 $\mu$m, and the cool dust that causes most of the
optical obscuration. The dark lanes at optical wavelengths result from
dust in the foreground of the optically emitting material, and
projection effects also may play a role.

\section{The most massive stars in the Antennae are not visible at optical wavelengths}

We now show that one of the inconspicuous regions at optical
wavelengths harbors the most massive stars in the Antennae.  For this
we use the imaging capability of the CVFs obtained with 6$''$ pixel
field of view, which offers a unique opportunity to study the
variation of spectral features from one area of the galaxy to the
other.

In Figure 1 we show as an example, a comparison between the CVF
spectra for the brightest 15 \mic\ peak in the overlap region and the
two nuclei. The distinct characteristic of this optically obscured
starburst knot is the [Ne\,III] line at 15.5 \mic\ above the
relatively enhanced continuum beyond 10 $\mu$m. Vigroux et al. (1996)
have shown that in the Antennae the 15 \mic\ continuum intensity is
well correlated with the [Ne~II] and the [Ne~III] line
intensities. This confirms that the enhancement of the continuum
$\geq$ 10 \mic\ is due to the thermal emission of hot dust heated by
the absorption of the ionizing photons emitted by young stars. This is
strengthened by the fact that the 15 \mic\ intensity increases
together with the [Ne~III]/[Ne~II] ratio (Vigroux et al. 1996).

For a given physical size, the [Ne~III]/[Ne~II] ratio is a measure of
the hardness of the UV flux. The [Ne\,III] to [Ne\,II] ratio is $\sim$
1 in the brightest area of the overlap region, and decreases to 0.1 or
below in the central regions of NGC\,4038 and in NGC\,4039 (Vigroux et
al. 1996). Using a [Ne~III]/[Ne~II] ratio of 1 and the diagnostic
arguments by Kunze et al. (1996), a single star equivalent effective
temperature of 44000 K is derived.  This would correspond to an O5
main sequence star with a mass of 50-60 M$_{\odot}$. A more adequate
treatment of star cluster evolution leads to a young cluster ($\sim$ 7
10$^6$ yr) starburst with IMF extending up to 100 M$_{\odot}$ (Kunze
et al. 1996). Therefore, using observations at optical wavelengths
only, could lead to biased low values for the high mass cut-off in the
IMF of interacting, luminous galaxies.

Multiwavelength observations of nearby starburst systems are
instructive when deriving the morphologies of galaxy populations at
high redshifts. Simulations that take into account band-shifting and
surface brightness dimming by Hibbard \& Vacca (1997) have shown that
nearby interacting systems that are luminous in the infrared, are the
best local analogs of the highest redshift galaxies found in the
Hubble Deep Field (HDF). Nevertheless, at z$\geq$1.5 the HDF is
sensitive to the rest-frame UV emission, and due to the presence of an
old population and/or dust it may be difficult to recover the global
morphology of the underlying systems (O'Connell, 1997). If, as in the
Antennae, the most intense starburst galaxies at high redshifts have
substantial amounts of dust (Franceschini et al 1994; Guiderdoni et
al. 1997), high sensitivity observations in the far-infrared and
submillimeter wavelength bands will be needed to reveal the true
global morphology of very distant galaxies.

\section{Conclusions}

1) The most intense starburst in the prototype merger NGC 4038/39
takes place in an off-nuclear region that is optically obscured. This
confirms the interpretation of the high far-infrared luminosity in
this system in terms of star formation.

2) The [Ne~III]/[Ne~II] emission line ratios in the mid-infrared
indicate that stars as massive as 60 M$_{\odot}$ are being formed in
that optically obscured region. Therefore, to derive the high mass end
of the Initial Mass Functions of the starbursts in luminous
interacting galaxies, observations at wavelengths longer than the
optical are needed.

3) The mid-infrared observations presented here suggest that about
15\% of the bolometric luminosity from NGC 4038/39 arises in an
off-nuclear region that is $\leq$ 100 pc in size.

4) The most prominent dust lanes in the optical image appear displaced
from the peak distribution of the warm dust and gas traced by the
mid-infrared emission. This could be due to different spatial
distributions of the warm and cool dust, and/or projection effects in
the optical appearance of dark lanes.

5) The infrared image also shows strong mid-infrared emission
associated with the optically prominent star forming ring in NGC 4038.

6) The effects of absorption by dust are even more dramatic in the
UV. Therefore, images of high redshift galaxies in their rest-frame UV
could lead to strong biases in the morphological classification, and
therefore in the scenarios of the history of galaxy formation.

\begin{acknowledgements} 
We thank Bradley Whitmore and an anonymous referee for useful
comments. This work was based on observations with ISO, an ESA project
with instruments funded by ESA Member States (especially the PI
countries: France, Germany, the Netherlands and the United Kingdom)
and with participation of ISAS and NASA. Based on observations made
with the NASA/ESA Hubble Space Telescope, obtained from data archive
at the Space Telescope Science Institute. STScI is operated by the
Association of Universities for Research in Astronomy, Inc. under the
NASA contract NAS 5-26555.
\end{acknowledgements}

%----------------------------------------------------------------------------

\vfill\eject

{\bf Figure caption}
\vskip .1in

The upper figure shows a superposition of the mid-infrared (12 -17
\mic\, red contours) image of the Antennae galaxies obtained with the
Infrared Space Observatory, on the composite optical image with V
(5252 \AA) and I (8269 \AA) filters recovered from the Hubble Space
Telescope archive. About half of the mid-infrared emission from the
gas and dust that is being heated by recently formed massive stars
comes from an off-nuclear region that is clearly displaced from the
most prominent dark lanes seen in the optical. The brightest
mid-infrared emission comes from a region that is relatively
inconspicuous at optical wavelengths. The ISOCAM image was made with a
1.5$''$ pixel field of view. Contours are 0.4, 1, 3, 5, 10, and 15
mJy.

The lower figure shows spectra of the brightest mid-infrared knot
(red) and of the nuclei of NGC 4038 (yellow) and NGC 4039 (black). The
rise of the continuum above 10 $\mu$m and strong NeIII line emission
observed in the brightest mid-infrared knot indicate that the most
massive stars in this system of interacting galaxies are being formed
in that optically obscured region, still enshrouded in large
quantities of gas and dust.

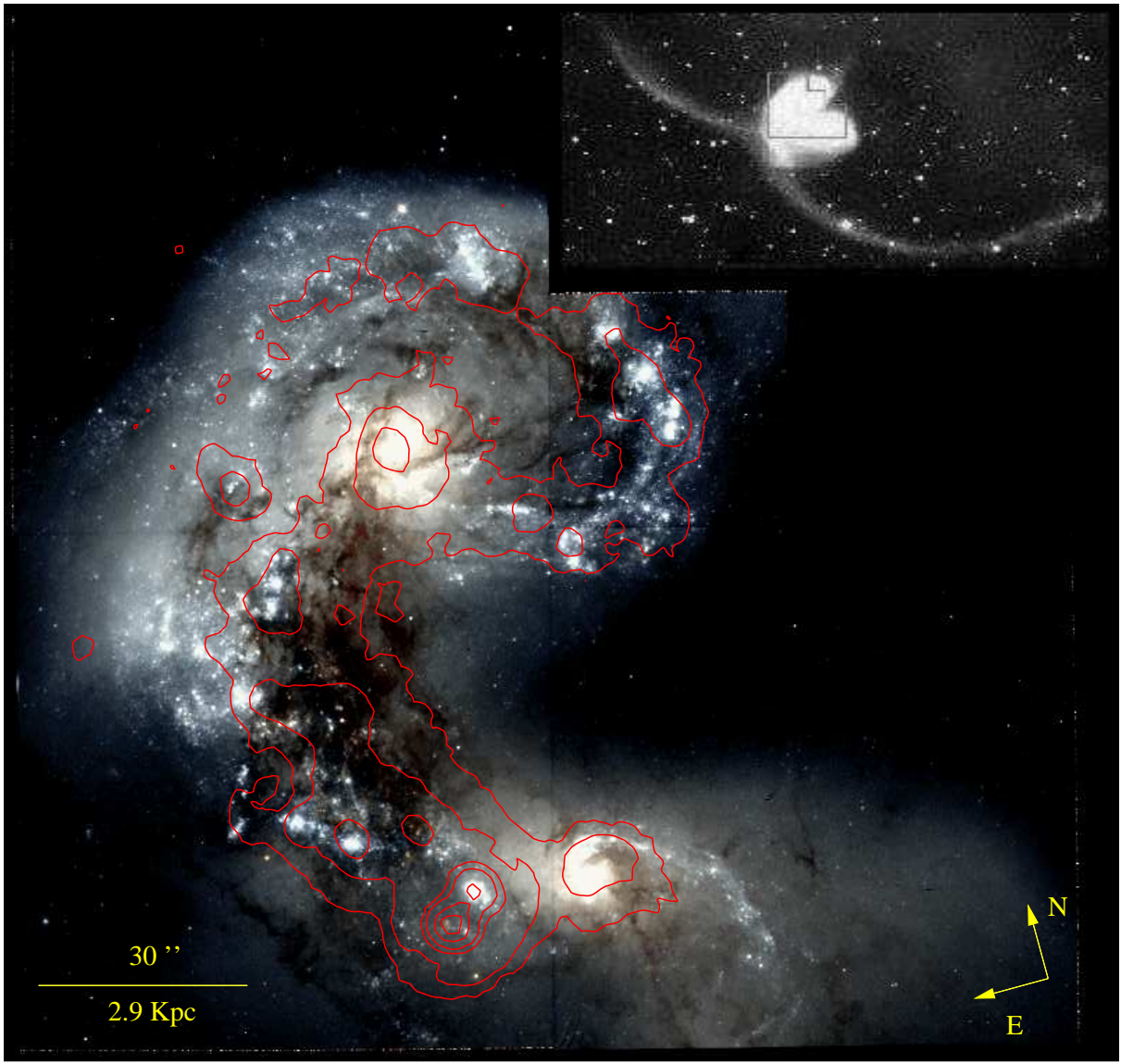
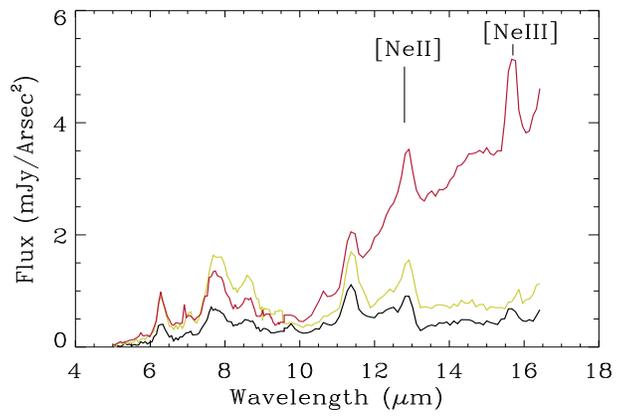
\end{document}